\documentstyle[prl,aps,epsf,floats]{revtex}
\begin{document}
\draft
\twocolumn[\hsize\textwidth\columnwidth\hsize\csname @twocolumnfalse\endcsname
\title{Band Kondo Effect 
       in the Doped 2D Hubbard Model
       }
\author{Bumsoo Kyung}
\address{Max Planck Institute for Physics of Complex Systems, 
         Noethnitzer Str. 38, 01187 Dresden, Germany} 
\date{23 February 1998}
\maketitle
\begin{abstract}

   We present strong numerical evidence for the 
band Kondo effect in the doped 2D Hubbard model by showing 
the systematic change of the density of states,
imaginary part of the self-energy, effective magnetic moment
and quasiparticle residue
upon decreasing the temperature.
Quadratically vanishing (in frequency)  
scattering rates near the Fermi 
energy and a linearly vanishing (in temperature) 
effective magnetic moment 
at low temperatures
strongly support
the screening 
of magnetic moments from singly occupied electrons by 
doped holes in the 2D Hubbard model.
\end{abstract}
\vskip2pc]
\narrowtext

   The Kondo effect\cite{Kondo:1964,Hewson:1993}
is realized by screening a local magnetic moment
by conduction electrons below a characteristic temperature
T$_{K}$.
Below this temperature conduction electrons form a singlet with
a localized electron at the impurity site and as a result
quench effectively magnetic degrees of freedom for a magnetic  
impurity.
Recently a possibility of the 
so called {\it band} Kondo effect  
in the finite dimensional Hubbard model
has received considerable attention,
since its discovery in the infinite dimensional Hubbard
model\cite{Jarrell:1992,Jarrell:1993,Georges:1996}.
In the latter limit, itinerant electrons or doped holes play a role of
conduction electrons and the other singly occupied electrons  
(due to a large Coulomb repulsion $U$) a role of
local magnetic impurities.
Consequently 
a sharp peak sensitive to the temperature
appears  
at the Fermi energy in the density of states.
In contrast to the usual Kondo or Anderson models%
\cite{Anderson:1961,Lee:1986},  
there is 
no clear distinction between conduction electrons in the Fermi sea
and localized 
electrons at the impurity sites in this situation.
The connection of this effect 
in the infinite dimensional Hubbard model  
with a Kondo-like screening 
is evident in that 
in this limit
the model is reduced to an effective single impurity 
Anderson model where 
a Kondo or Abrikosov-Suhl resonance  
has been 
well established.
Because of the absence of a formal connection in the finite dimensional
model, however,  
it is quite uncertain whether the band Kondo effect 
survives or not in this case.

   Recently 
this problem was studied by Bulut {\it et al.}\cite{Bulut:1994}
by means of quantum Monte Carlo (QMC) calculations
for the 2D Hubbard model. 
Although these authors showed a developing
peak near the Fermi energy in the density of states by decreasing
the temperature,
it is not clear whether this peak is associated with the onset of 
the Fermi liquid behavior or not,
and more importantly whether it is due to 
the screening 
of magnetic moments (from singly occupied electrons) by 
doped holes.
In this Letter,  
we will present strong numerical evidence for the 
band Kondo effect in the doped 2D Hubbard model by showing 
the systematic change of the density of states,
imaginary part of the self-energy, effective magnetic moment
and quasiparticle residue
upon decreasing the temperature.

   There are fundamental differences between the two and infinite 
dimensional Hubbard models. 
In contrast to the latter model where
an Abrikosov-Suhl resonance appears 
for any infinitesimally small doping of a half-filled band, 
strong 2D spin fluctuations
near half-filling\cite{Kyung:19971,Vilk:1997}
easily
destroy the singlet formation in the 2D model.
Hence, 
forming a quasiparticle state at the Fermi energy,
if it is possible,
is expected to happen only below a certain electron concentration in the 
2D Hubbard model.
In order to take into account  
strong 2D critical spin fluctuations properly, 
we impose the following three exact sumrules to 
the spin, charge, and particle-particle susceptibilities%
\cite{Kyung:19971,Vilk:1997}:
\begin{eqnarray}
\frac{T}{N}\sum_{q}\chi_{sp}(q) & = & n-2\langle n_{\uparrow}n_{\downarrow}
                                         \rangle
                                                         \nonumber  \\
\frac{T}{N}\sum_{q}\chi_{ch}(q) & = & n+2\langle n_{\uparrow}n_{\downarrow}
                                         \rangle-n^{2}
                                                         \nonumber  \\
\frac{T}{N}\sum_{q}\chi_{pp}(q) & = & \langle n_{\uparrow}n_{\downarrow}
                                         \rangle
                                               \; .
                                                           \label{eq1}
\end{eqnarray}
$T$ and $N$ are the absolute temperature and 
number of lattice sites. 
$q$ is a compact notation for $(\vec{q},i\nu_{n})$ where
$i\nu_{n}$ are either Fermionic or 
Bosonic Matsubara frequencies. 
The dynamical spin, charge and particle-particle susceptibilities are 
calculated by 
\begin{eqnarray}
\chi_{sp}(q)&=&\frac{2\chi^{0}_{ph}(q)}{1-U_{sp}\chi^{0}_{ph}(q)}
                                                         \nonumber  \\
\chi_{ch}(q)&=&\frac{2\chi^{0}_{ph}(q)}{1+U_{ch}\chi^{0}_{ph}(q)}
                                                         \nonumber  \\
\chi_{pp}(q)&=&\frac{ \chi^{0}_{pp}(q)}{1+U_{pp}\chi^{0}_{pp}(q)}
                                               \; .
                                                           \label{eq2}
\end{eqnarray}
$\chi^{0}_{ph}(q)$ and 
$\chi^{0}_{pp}(q)$ are 
irreducible particle-hole and particle-particle susceptibilities, 
respectively, which are computed from
\begin{eqnarray}
\chi^{0}_{ph}(q) & = & - \frac{T}{N}\sum_{k}G^{0}(k-q)G^{0}(k)
                                                         \nonumber  \\
\chi^{0}_{pp}(q) & = &  \frac{T}{N}\sum_{k}G^{0}(q-k)G^{0}(k)
                                               \; ,
                                                           \label{eq3}
\end{eqnarray}
where 
$G^{0}(k)$ is the noninteracting Green's function.
$U_{sp}$, $U_{ch}$, and $U_{pp}$ in Eq.~\ref{eq2} are    
renormalized interaction constants for each channel which are 
calculated self-consistently 
by making an ansatz
$U_{sp} \equiv U\langle n_{\uparrow}n_{\downarrow} \rangle/
(\langle n_{\uparrow} \rangle
\langle n_{\downarrow} \rangle)$\cite{Vilk:1997}
in Eq.~\ref{eq1}.
By defining 
$U_{sp}$, $U_{ch}$, and $U_{pp}$    
this way, the Mermin-Wagner theorem\cite{Mermin:1966}
as well as correct atomic limit for large $\omega$
are satisfied 
simultaneously\cite{Kyung:19971}. 
In order to find the chemical potential for interacting electrons,
first
we calculate
Eqs.~(\ref{eq1})-(\ref{eq3}) and the self-energy 
(Eqs.~(\ref{eq1}) in Ref.~\onlinecite{Kyung:19971})
with the noninteracting Green's function whose 
noninteracting chemical potential gives 
a desired electron concentration. 
Then, 
the chemical potential for interacting electrons is determined 
in such a way that the calculated electron concentration 
with the interacting Green's function 
is the same as the desired value.
Throughout the calculations the unit of energy is $t$ and all energies 
are measured from the chemical potential $\mu$. 
We used a $128 \times 128$ lattice in momentum space 
and performed the calculations by means of 
well-established fast Fourier transforms
(FFT).
It should be also noted that we used a real frequency formulation
in Eqs.~(\ref{eq1})-(\ref{eq3}) to avoid any possible uncertainties
associated with numerical analytical continuation.

\begin{figure}
 \vbox to 7.5cm {\vss\hbox to -5.0cm
 {\hss\
       {\includegraphics{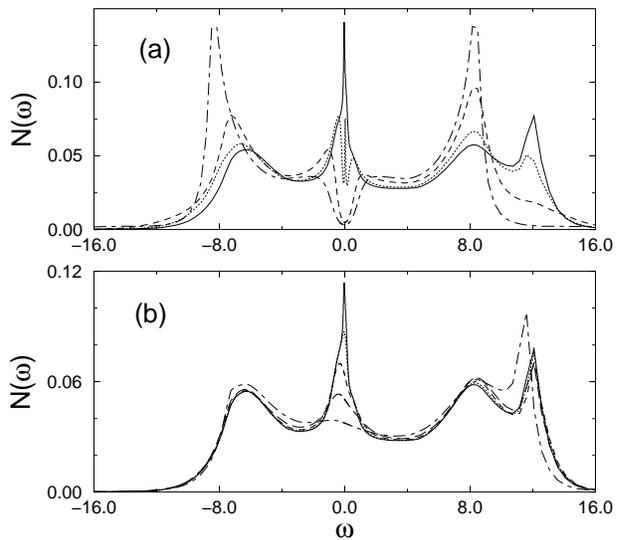}
       }
  \hss}
 }
\caption{Density of states for $U=8$ 
         (a) at $n=1.0$, 0.91, 0.85 and 0.80 denoted as 
             the dot-dashed, 
             dashed, dotted, and solid curves, respectively and 
         (b) at $T=0.8$, 0.4, 0.2, 0.1 and 0.05 denoted as 
             the dot-dashed, long-dashed,
             dashed, dotted, and solid curves, respectively  
             at $n=0.80$.
             In (a) for $n=1.0$ and 0.91, $T$ is 0.05, while
             for $n=0.85$ and 0.80, $T$ is 0.005.} 
\label{fig1}
\end{figure}

   We start in Fig.~\ref{fig1} by studying 
the density of states at various electron
concentrations
$n=1.0$, 0.91, 0.85 and 0.80  
(Fig.~\ref{fig1}(a)) and  
at $n=0.80$ for various temperatures (Fig.~\ref{fig1}(b))  
for $U=8$.
Within our numerical calculations, we do not have the self-consistent 
solution for $U_{sp}$ at lower temperatures than 0.05 near half-filling.
Note that the spin-spin correlation length exponentially 
increases like
$\xi \sim exp(constant/T)$ at low temperatures near half-filling.
At $n=1.0$ (dot-dashed curve), strong 2D critical spin fluctuations due to 
the Mermin-Wagner theorem deplete the spectral weight near the 
Fermi energy, leading to an antiferromagnetic pseudogap. 
Since the imaginary part of the self-energy is proportional to 
$\xi$ 
at $\omega=0$ for low temperatures near half-filling%
\cite{Kyung:19971,Vilk:1997}, 
scattering rates increase exponentially with decreasing temperature, 
eventually leading to an antiferromagnetic insulator at zero 
temperature.
At $n=0.91$ (dashed curve), strong spin fluctuations are still 
dominating near the Fermi energy as the strong persistence of  
the antiferromagnetic pseudogap 
indicates.
Decreasing the temperature still gives rise to exponentially
increasing scattering rates. 
This is in clear contrast to the infinite dimensional model where
a sharp quasiparticle peak associated with
an Abrikosov-Suhl resonance appears by infinitesimally small 
doping of a half-filled band.
At $n=0.85$ (dotted curve),
most of the antiferromagnetic pseudogap is closed and a small 
peak develops at the Fermi energy. 
A careful investigation of the imaginary part of the self-energy
(not shown in this Letter),
however, still shows a small remnant of the 2D critical 
fluctuations near the Fermi energy.
At $n=0.80$ (solid curve), 
the antiferromagnetic pseudogap is completely closed by doping and 
a narrow, sharp spectral weight appears at the Fermi energy.

\begin{figure}
 \vbox to 7.5cm {\vss\hbox to -5.0cm
 {\hss\
       {\includegraphics{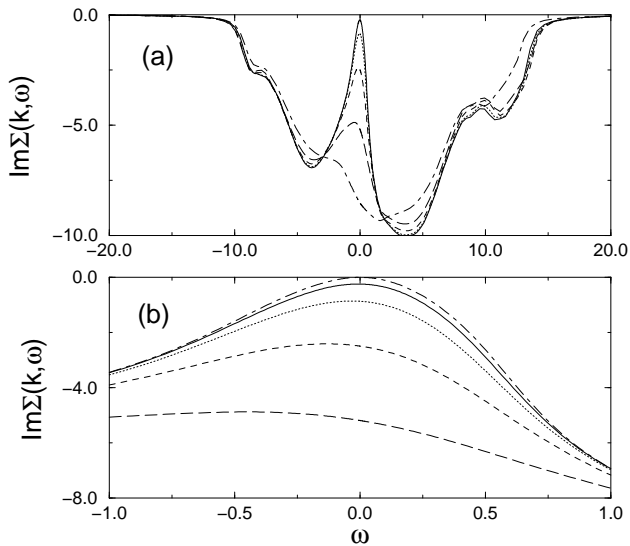}
       }
  \hss}
 }
\caption{Imaginary part of the self-energy 
             for $U=8$ and $n=0.80$ 
             at the Fermi surface 
         (a) for $T=0.8$, 0.4, 0.2, 0.1 and 0.05 denoted as 
             the dot-dashed, long-dashed,
             dashed, dotted, and solid curves, respectively  
             in the wide range of frequency axis and 
         (b) for $T=0.4$, 0.2, 0.1 and 0.05 denoted as 
             the long-dashed,
             dashed, dotted, and solid curves, respectively  
             in the low frequency regime.
             The dot-dashed curve in (b) is for $T=0.005$.}
\label{fig2}
\end{figure}

   In Fig.~\ref{fig1}(b), the systematic change of the density of 
states for $U=8$ and $n=0.80$ is presented for 
$T=0.8$ to 0.05.  
As the temperature is decreased,  
the general shape of the 
density of states remains unchanged except near the Fermi energy.
A small spectral weight in the intermediate frequency regime is transferred
near the Fermi energy and 
a featureless background at high temperatures 
is transformed into a sharp peak at the Fermi energy with decreasing 
temperature.
In order to find that this feature is associated with
the onset of the Fermi liquid, we present the imaginary part of the 
self-energy 
in Fig.~\ref{fig2} for the same parameters as in 
Fig.~\ref{fig1}(b).

   First 
$Im \Sigma(\vec{k},\omega)$
is shown in the wide range of frequency axis in Fig.~\ref{fig2}(a).
With decreasing temperature scattering rates 
are progressively decreased particularly near the  
Fermi energy. 
The low frequency behavior of the scattering rates
is shown in Fig.~\ref{fig2}(b).
Below $T=0.1-0.2$, the imaginary part of the self-energy vanishes
quadratically near the Fermi energy with a small constant shift
due to a finite temperature effect.
The same quantity for $T=0.005$
is also plotted as 
the dot-dashed curve where
the quadratically vanishing scattering rates 
are more visible.  
The log-log plot of the scattering rates vs. frequency
for $T=0.005$ shows
$Im \Sigma(\vec{k}_{F},\omega) \sim \omega^{1.94}$
in an interval
of $\pm [0.06-0.61]$.

\begin{figure}
 \vbox to 7.5cm {\vss\hbox to -5.0cm
 {\hss\
       {\includegraphics{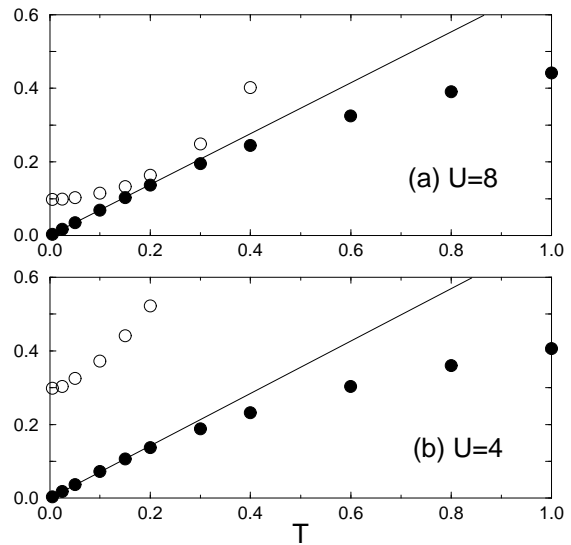}
       }
  \hss}
 }
\caption{Effective magnetic moment (filled circles) and 
         quasiparticle residue (open circles) 
         (a) for $U=8$, $n=0.80$ and      
         (b) for $U=4$, $n=0.87$.
         The solid lines are a linear interpolation of the 
         effective magnetic moment at low temperatures.}
\label{fig3}
\end{figure}

   In order to establish a more firm ground that this is indeed
due to the screening 
of magnetic moments by doped holes,  
we show the effective magnetic moment defined as 
$T\chi_{sp}(0,0)$ upon decreasing the temperature 
(filled circles in Fig.~\ref{fig3}(a)).
Below $T=0.2$, the effective magnetic moment vanishes linearly 
in temperature, a clear indication of 
the Kondo screening.
Above $T=0.2$, the effective magnetic moment  
deviates significantly from a straight line 
and appears gradually saturating at high temperatures.
The quasiparticle residue is also presented as the open circles
in Fig.~\ref{fig3}(a).
With decreasing temperature it saturates at 0.1 around 
$T=0.05$.
Hence, the effective mass of the 
quasiparticle becomes ten times heavier than the bare electron mass
below $T=0.05$.  

\begin{figure}
 \vbox to 7.5cm {\vss\hbox to -5.0cm
 {\hss\
       {\includegraphics{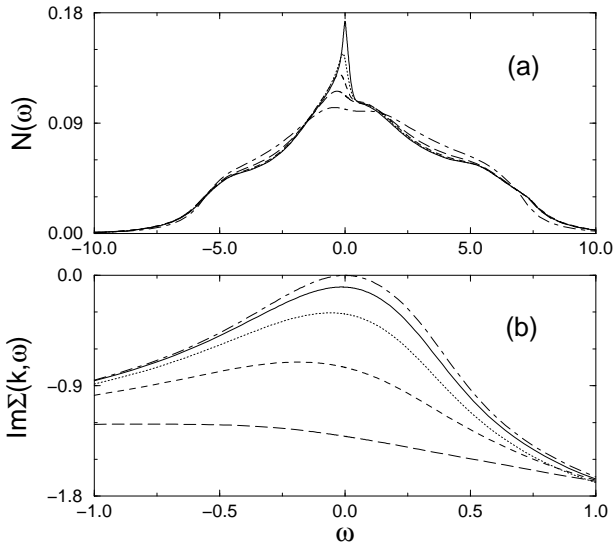}
       }
  \hss}
 }
\caption{(a) Density of states for
             $T=0.8$, 0.4, 0.2, 0.1 and 0.05 denoted as 
             the dot-dashed, long-dashed,
             dashed, dotted, and solid curves, respectively  
             and
         (b) imaginary part of the self-energy 
             at the Fermi surface
             for $T=0.4$, 0.2, 0.1 and 0.05 denoted as 
             the long-dashed,
             dashed, dotted, and solid curves, respectively  
             in the low frequency regime
             for $U=4$ and $n=0.87$. 
             The dot-dashed curve in (b) is for $T=0.005$.}
\label{fig4}
\end{figure}

   We also performed the same calculations for $U=4$.
Due to a smaller coupling strength, the complete destruction of 
the 2D critical spin fluctuations happens at $n=0.87$.
In Fig.~\ref{fig4}(a) the density of 
states for $U=4$ and $n=0.87$ is shown at
$T=0.8$ to 0.05.  
The general shape of the density of states for $U=4$ is 
quite different from that for $U=8$ but instead closer to 
that of noninteracting electrons.
With decreasing temperature some spectral weight in the 
intermediate frequency regime is transferred near the Fermi
energy to build a sharp peak which can be interpreted as a 
Kondo resonance.
We present the scattering rates 
in the low frequency regime 
in Fig.~\ref{fig4}(b).  
As in Fig.~\ref{fig2}(b),
below $T=0.1-0.2$, the imaginary part of the self-energy vanishes
quadratically near the Fermi energy with a small constant shift.
The same quantity for $T=0.005$ is also presented as
the dot-dashed curve.  
The log-log plot of the scattering rates vs. frequency
for $T=0.005$ shows
$Im \Sigma(\vec{k}_{F},\omega) \sim \omega^{1.95}$
in an interval
of $\pm [0.05-0.37]$.
In Fig.~\ref{fig3}(b), we show 
the effective magnetic moment (filled circles) and quasiparticle 
residue (open circles) for $U=4$.
The effective magnetic moment for $U=4$ also vanishes linearly in temperature
below $T=0.2$.
The quasiparticle residue saturates at 0.29  
below $T=0.025$.

\begin{figure}
 \vbox to 7.5cm {\vss\hbox to -5.0cm
 {\hss\
       {\includegraphics{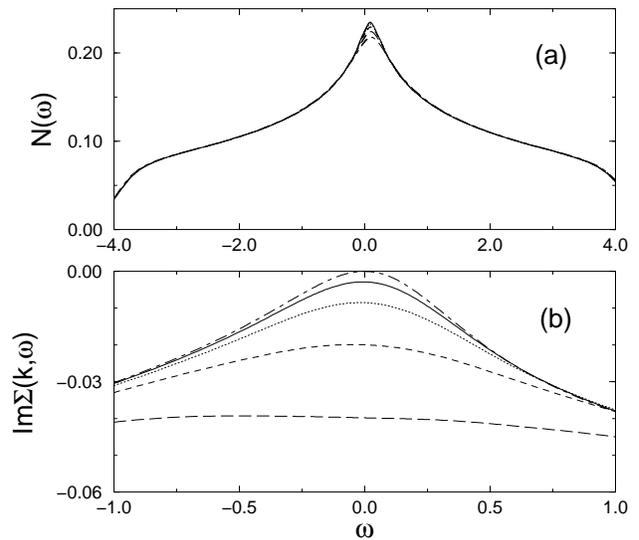}
       }
  \hss}
 }
\caption{(a) Density of states for
             $T=0.8$, 0.4, 0.2, 0.1 and 0.05 denoted as 
             the dot-dashed, long-dashed,
             dashed, dotted, and solid curves, respectively  
             and
         (b) imaginary part of the self-energy 
             at the Fermi surface
             for $T=0.4$, 0.2, 0.1 and 0.05 denoted as 
             the long-dashed,
             dashed, dotted, and solid curves, respectively  
             in the low frequency regime
             for $U=1$ and $n=0.95$. 
             The dot-dashed curve in (b) is for $T=0.005$.}
\label{fig5}
\end{figure}

   Finally we show in Fig.~\ref{fig5} 
the density of states and imaginary part of 
the self-energy for $U=1$ and $n=0.95$ 
where critical spin fluctuations are completely destroyed. 
For this weak coupling strength, 
magnetic moments from singlely occupied electrons are not  
well developed and thus the screening 
of the magnetic moments by doped holes  
is questionable.
The density of states from $T=0.8$ to 0.05  
in Fig.~\ref{fig5}(a)
is very close to 
that for noninteracting electrons with a small constant shift in the 
frequency axis.
Within our numerical resolution it is difficult to find any 
developing peak near the Fermi energy 
by decreasing the temperature.
The imaginary part of the self-energy  
in Fig.~\ref{fig5}(b), however, clearly exhibits the drastic 
change of the scattering rates to the quadratic behavior near 
the Fermi energy with decreasing temperature.
The log-log plot of the scattering rates vs. frequency
for $T=0.005$ shows
$Im \Sigma(\vec{k}_{F},\omega) \sim \omega^{1.97}$
in an interval
of $\pm [0.04-0.18]$.
The quasiparticle residue is found to saturate at 0.92
and the effective magnetic moment also vanishes linearly in temperature
with decreasing temperature.
All these findings for $U=1$ 
are also consistent with the existence of the 
Kondo screening in the doped 2D Hubbard model  
found earlier 
for larger interaction 
strengths.

   In summary,
we have presented strong numerical evidence for the 
band Kondo effect in the doped 2D Hubbard model by showinging 
the systematic change of the density of states,
imaginary part of the self-energy, effective magnetic moment
and quasiparticle residue
upon decreasing the temperature.
Quadratically vanishing (in frequency)  
scattering rates near the Fermi 
energy and a linearly vanishing (in temperature) 
effective magnetic moment 
at low temperatures
strongly support
the screening 
of the magnetic moments from singly occupied electrons by 
doped holes in the doped 2D Hubbard model.
Our numerical calculations indicate that 
the band Kondo effect in the doped 2D Hubbard
model persists in the strong to weak coupling 
regimes.

    The author would like to thank Prof. P. Fulde,  
and Drs. S. Blawid, R. Bulla, 
M. Laad and numerous other colleagues 
in the Max Planck Institute
for Physics of Complex Systems 
for useful discussions.   
\end{document}